\documentclass{jkas}

\def\year{2020} 
\def\month{June} 
\def\volume{53} 
\def\issue{3} 
\def\beginpage{1} 
\def\endpage{9} 
\def\received{March 5, 2020} 
\def\accepted{April 15, 2020} 
\date{Received \received; accepted \accepted}

\def\ie{{i.e.,\ }}
\def\eg{{e.g.,\ }}
\def\kms{~{\rm km~s^{-1}}}
\def\cm3{~{\rm cm^{-3}}}

\def\muG{~{\mu\rm G}}

\usepackage{flushend}

\title{Semi-analytic Models for Electron Acceleration in Weak ICM Shocks}

\author{Hyesung Kang}

\affil{Department of Earth Sciences, Pusan National University, Busan 46241, Korea; \email{hskang@pusan.ac.kr}}

\def\corrauthor{H. Kang}
\def\runningauthor{Kang}
\def\runningtitle{Electron Acceleration in Weak ICM Shocks}
\def\keywords{acceleration of particles --- cosmic rays ---
galaxies: clusters: general --- shock waves}

\def\abstracttext{
We propose semi-analytic models for the electron momentum distribution in weak shocks that accounts for 
both {\it in situ} acceleration and re-acceleration through diffusive shock acceleration (DSA).
In the former case, a small fraction of incoming electrons are assumed to be reflected at the shock ramp 
and pre-accelerated to the so-called injection momentum, $p_{\rm inj}$, above which particles can
 diffuse across the shock transition and participate in the DSA process.
This leads to the DSA power-law distribution extending from the smallest momentum of reflected electrons, $p_{\rm ref}$, all the way to the cutoff momentum, $p_{\rm eq}$, constrained by radiative cooling.
In the later case, fossil electrons, specified by a power-law spectrum with a cutoff, are assumed to 
be re-accelerated also from $p_{\rm ref}$ up to $p_{\rm eq}$ via DSA.
We then show that, in the {\it in situ} acceleration model, the amplitude of radio synchrotron emission
depends strongly on the shock Mach number, whereas it varies rather weakly in the re-acceleration model.
Considering rather turbulent nature of shocks in the intracluster medium, such extreme dependence for the 
{\it in situ} acceleration might not be compatible with relatively smooth surface brightness 
of observed radio relics.}
\begin{document}
\jkashead 
\section{Introduction}

Cosmological hydrodynamic simulations predicted that the intracluster medium (ICM) could encounter shocks several times
on average during the formation 
of the large scale structures in the Universe \citep[e.g.,][]{ryu03,vazza09}.
As in the case of astrophysical shocks such as the Earth's bow shock and supernova remnants, ICM shocks are expected to
produce cosmic-ray protons (CRp) and electrons (CRe) via diffusive shock acceleration (DSA) \citep[e.g.][]{bell78, dru1983,brunetti2014}.
Many of merger-driven shocks have been observed and identified as ``radio relic shocks'' in the outskirts of galaxy clusters 
through radio synchrotron radiation from shock-accelerated CRe with
Lorentz factor $\gamma_e\sim 10^3-10^4$ \citep[e.g.][]{vanweeren10, vanweeren19, kang12}.
Shocks formed in the hot ICM are weak with the sonic Mach number, $M_s \lesssim 4$, which
can be inferred from the observed radio spectral index, $\alpha_{\nu}= {{(M_s^2+3)}/{2(M_s^2-1)}}$, using the test-particle prediction of DSA \citep[e.g.][]{kang11, kang16}. 

In this discussion, the shock is specified by the sonic Mach number, $M_{\rm s}$, and preshock temperature, $T_1$,
where the subscripts, $1$ and $2$, denote the preshock and postshock states, respectively.
The momentum distribution, $f(p)$, scales with the upstream gas density, $n_1$, and so it does not need to be specified.
For quantities related with synchrotron emission and cooling, the preshock magnetic field strength, $B_1=1 \muG$, is adopted.
The plasma beta refers the ratio of thermal to magnetic pressures, $\beta = P_{\rm gas}/P_{\rm B}$, in the background ICM.
Common symbols in physics are used: \eg $m_{\rm e}$ for the electron mass, $m_{\rm p}$ for the proton mass, $c$ for the speed of light, 
and $k_{\rm B}$ for the Boltzmann constant.

Suprathermal particles above the so-called {\it injection momentum}, $p_{\rm inj}$, have gyroradii long enough to diffuse across the shock transition and may
participate in DSA, {\it aka} Fermi 1st-order acceleration, if scattering MHD/plasma waves of sufficient amplitudes are present \citep[e.g.,][]{dru1983}.
The pre-acceleration of thermal particles to $p_{\rm inj}$, \ie the `injection problem',
has been a longstanding, key problem in the DSA theory \citep[e.g.,][]{maldru01,kjg02,marcowith2016}.
According to plasma simulations of {\it quasi-parallel} shocks \citep{caprioli14,caprioli15, ha2018}, 
some of incoming protons are specularly reflected by the overshoot in the shock potential and undergo shock drift acceleration (SDA) at the shock front, resulting in the self-excitation of upstream waves via
both resonant and non-resonant streaming instabilities.
Then the protons are scattered around the shock by those waves, which leads to the formation of the DSA power-law spectrum above
$p_{\rm inj}\sim (3.0-3.5) p_{\rm th,p}$, where
$p_{\rm th,p}=(2m_p k_B T_{2})^{1/2}$ is the postshock thermal proton momentum.

On the other hand, electrons are known to be injected and accelerated preferentially at {\it quasi-perpendicular} shocks,
which involves kinetic processes on electron kinetic scales much smaller than ion scales \citep[e.g.,][]{balogh13}.
Earlier studies on the electron pre-acceleration via self-generated waves focused mainly on high Mach number shocks 
in $\beta\sim1$ plasma, which are relevant for supernova blast waves
\citep[e.g.,][]{levinson1992, levinson1996,amano09,riqu11}.
\citet{guo14} showed, through particle-in-cell (PIC) simulations, that 
in weak quasi-perpendicular shocks in high beta ICM plasma, 
a small fraction of incoming electrons are reflected due to the magnetic mirror and energized via SDA,
while the backstreaming electrons excite oblique waves by the electron firehose instability (EFI). 
Due to scattering of electrons between the shock ramp and EFI-induced waves in the shock foot, 
the pre-accelerated electrons seem to form a DSA power-law spectrum through a Fermi I-like acceleration.
However, \citet{kang2019} showed that such pre-acceleration is effective only in supercritical shocks with $M_{\rm s} \gtrsim 2.3$.
Moreover, they suggested that suprathermal electrons may not be energized all the way to $p_{\rm inj} \sim 150 p_{\rm th,e}$
(where $p_{\rm th,e}=(2m_e k_B T_{2})^{1/2}$), because the growth of longer waves via the EFI is saturated.
On the other hand, \citet{trotta2019} and \citet{kobzar2019} have demonstrated through hybrid simulations with test-particle electrons ($\beta\approx 1$)
and PIC simulations ($\beta\approx 5$), respectively, that at supercritical quasi-perpendicular shocks
the rippling of shock surface excited by Alfv\'en Ion Cyclotron (AIC) instability could induce multi-scale fluctuations, 
leading to the pre-acceleration of electrons beyond $p_{\rm inj}$.
Whereas the critical Mach number above which the shock rippling becomes active was estimated to be $M_{\rm A,crit}\approx 3.5$ for $\beta\approx 1$ shocks
\citep{trotta2019}, this problem needs to be investigated for higher $\beta$ shocks.

Although the DSA model seems to provide a simple and natural explanation for some observed properties of radio relics,
such as thin elongated shapes, postshock spectral steepening due to aging electron population, and polarization vectors indicating
perpendicular magnetic field directions, there remain some unresolved problems that need further investigation.
First of all, the pre-acceleration of thermal electrons to suprathermal energies and the subsequent injection
into the DSA process still remains rather uncertain, especially at subcritical shocks
with $M_s\lesssim 2.3$ \citep{kang2019}.
Secondly, the fraction of observed merging clusters with detected radio relics is only $\sim 10$ \% \citep{feretti12}, 
while numerous quasi-perpendicular shocks are expected to form in the ICM \citep{wittor17,roh2019}.
Thirdly, in a few cases, the sonic Mach number inferred from X-ray observations 
is smaller than that estimated from radio spectral index of radio relics,
\ie $M_{\rm X} < M_{\rm radio}$ \citep{akamatsu13,kang16}.
Thus re-acceleration of fossil CRe, pre-existing in the ICM, has been suggested as a possible resolution for these puzzles that
the DSA model with `{\it in situ} injection only' leaves unanswered
\citep[e.g.,][]{kang12,kang2017, kang16}.

Based on what we have learned from the previous studies, here we propose semi-analytic models for 
the momentum distribution function of CRe, $f(p)$, in the two scenarios of DSA at weak quasi-perpendicular shocks in the test-particle regime:
(1) {\it in situ} acceleration model in which electrons are injected directly from the background thermal pool at the shock,  and 
(2) re-acceleration model in which pre-existing fossil CRe are accelerated.
Although it remains largely unknown if and how CRe are accelerated at subcritical shocks, 
in this paper we take a heuristic approach and assume that DSA operates at shocks of all Mach numbers.

In the next section we describe in details the semi-analytic models for $f(p)$ along with in-depth discussion on
underlying physical justification.
In Section \ref{s3} we demonstrate how our model can be applied to weak shocks in the ICM and discuss
observational implications.
A brief summary will be given in Section \ref{s4}.

\section{Semi-Analytic DSA Model}
\label{s1}

The physics of collisionless shocks depends on various shock parameters including the sonic Mach number, $M_{\rm s}$, the plasma beta, $\beta$, and the obliquity angle, $\theta_{\rm Bn}$, between the upstream background magnetic field direction and the shock normal \citep[e.g.,][]{balogh13}.
For instance, collisionless shocks can be classified as {\it quasi-parallel} ($Q_\parallel$, hereafter) shocks with $\theta_{\rm Bn}\lesssim 45^{\circ}$ and {\it quasi-perpendicular} ($Q_\perp$, hereafter) shocks with $\theta_{\rm Bn}\gtrsim 45^{\circ}$. 
CRp are known to be accelerated efficiently at $Q_\parallel$-shocks, while CRe are accelerated preferentially 
at $Q_\perp$-shocks \citep{gosling1989,burgess2007,caprioli14, guo14}.

In this study, we focus on the electron acceleration at $Q_\perp$-shocks with $M_{\rm s} \lesssim 4$ that are expected to form in the ICM. 
Most of the kinetic problems involved in the electron acceleration, including the shock criticality, excitation of waves via microinstabilites, and wave-particle interactions, have been investigated
previously for shocks in $\beta\sim1$ plasma such as the solar wind and the interstellar medium
\citep[see][]{balogh13,marcowith2016}.
Although a few studies, using kinetic PIC simulations, have recently considered weak shocks in the high $\beta$ ICM environment \citep{guo14, matsukiyo15, kang2019, kobzar2019},
full understanding of the electron injection and acceleration into the regime of genuinely diffusive scattering has yet to come.

The main difficulty in reaching such a goal is the severe computational requirements to perform PIC simulations for high $\beta$ shocks;
the ratio of the proton Larmor radius to the electron skip depth increases with $\beta^{1/2}$.
Moreover, to properly study these problems, PIC simulations in at least two-dimensional domains
extending up to several proton Larmor radii are required, 
because kinetic instabilities induced by both protons and electrons may excite waves on multi-scales that
propagate in the direction oblique to the background magnetic fields.

\subsection{Particle Injection to DSA}
\label{s2.1}

In this section, we review the current understandings of the injection problem that have been obtained previously through plasma hybrid and PIC simulations.
Suprathermal particles, both protons and electrons, with $p \gtrsim 3 p_{\rm th,p}$ could 
diffuse across the shock both upstream and downstream, and participate in the DSA process,
because the shock thickness is of the order of the gyroradius of postshock thermal protons.
Thus the injection momentum is often parameterized as
\begin{equation}
p_{\rm inj} = Q_{\rm i,p} \cdot p_{\rm th,p},
\end{equation}
where the injection parameter is estimated to range $Q_{\rm i,p}\sim 3.0-3.5$,
according to the hybrid simulations of $Q_\parallel$ shocks in $\beta\sim 1$ plasma \citep{caprioli14, caprioli15,ha2018}.
On the other hand, \citet{ryu2019} showed that the DSA power-law with $Q_{\rm i,p}\approx 3.8$ gives the postshock 
CRp energy density less than 10 \% of the shock kinetic energy density for $M_{\rm s} \lesssim 4$, 
\ie $E_{\rm CRp} < 0.1 E_{\rm sh}$ (where $E_{\rm sh}=\rho_1 u_s^2/2$).

The electron injection at $Q_\perp$-shocks involves somewhat different processes, which can be summarized as follows:
(1) the reflection of some of incoming electrons at the shock ramp due to magnetic deflection, leading to a beam of backstreaming electrons,
(2) the energy gain from the motional electric field in the upstream region through shock drift acceleration (SDA),
(3) the trapping of electrons near the shock due to the scattering by the upstream waves, which are excited by backstreaming electrons via the EFI,
and (4) the formation of a suprathermal tail for $p\gtrsim p_{\rm ref}$ with a power-law spectrum, which seems consistent with the test-particle DSA prediction \citep{guo14,matsukiyo15}.
Here, $p_{\rm ref}$ represents the lowest momentum of the reflected electrons above which the suprathermal power-law tail develops. 
This is again parameterized as
\begin{equation}
p_{\rm ref}= Q_{\rm i,e} \cdot p_{\rm th,e},
\end{equation}
with the injection parameter, which is assumed to range $Q_{\rm i,e}\sim 3.5-3.8$ as in the case of 
$p_{\rm inj}$ \citep[e.g.,][]{guo14,kang2019}.

Recently, \citet{kang2019} showed that the electron pre-acceleration through the combination of reflection, SDA, and EFI may operate only in {\it supercritical} $Q_\perp$-shocks with $M_{\rm s} \gtrsim 2.3$ in $\beta\sim 100$ plasma.
In addition, they argued that the EFI alone may not energize the electrons all the way to $p_{\rm inj}$, unless there are pre-existing turbulent waves with wavelengths longer than those of the EFI-driven waves.
As mentioned earlier, on the other hand, \citet{trotta2019} and \citet{kobzar2019} showed through 2D simulations 
that the suprathermal tail may extend to beyond $p_{\rm inj}$ in the presence of multi-scale turbulence excited by the shock rippling instability.
But \citet{trotta2019} suggested that the critical Mach number, at which the shock surface rippling
starts to develop, is $M_{\rm s} \approx 3.5$ 
in $\beta\approx 1$ plasma.
Hence, we still need to answer the following questions in future studies:
(1) if and how the electron injection occurs at subcritical shocks with $M_{\rm s} \lesssim 2.3$, and
(2) how the critical Mach number for the shock surface rippling varies with shock parameters such as $\beta$ and $\theta_{\rm Bn}$.

On the other hand, X-ray and radio observations of several radio relics indicate the efficient electron acceleration even at subcritical shocks with $1.5 \lesssim M_{\rm s} \lesssim 2.3$ \citep{vanweeren19}.
Hence, in the discussion below, we heuristically assume that the DSA power-law spectrum of the accelerated electrons, $f_{\rm e,inj}\propto p^{-q}$ (where $q = 4M_{\rm s}^2/(M_{\rm s}^2-1)$),
develops from $\sim p_{\rm ref}$ all the way to the cutoff momentum $p_{\rm eq}$ (see below) at $Q_\perp$-shocks of all Mach numbers.
This hypothesis needs to be examined in future studies for high $\beta$ shocks.

Nevertheless, we refer Figure 4 of \cite{park2015}, in which the acceleration of both protons and electrons at strong $Q_\parallel$-shocks ($M_s=40$) were investigated through 1D PIC simulations.
There, electrons form a DSA power-law for $p\gtrsim p_{\rm pref}$, because local fields become
quasi-perpendicular at some parts of the shock surface due to turbulent magnetic field amplification
driven by the strong non-resonant Bell instability.

\begin{figure}[t]
\vskip -0.5cm
\centerline{\includegraphics[width=0.50\textwidth]{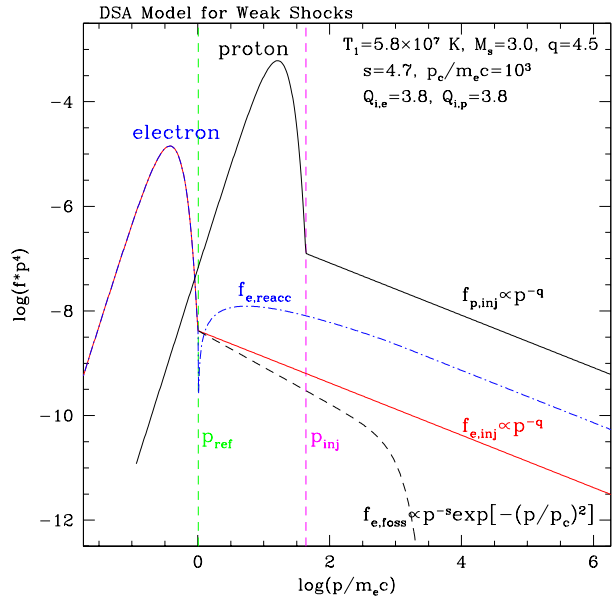}}
\vskip -0.5cm
\caption{
Semi-analytic functions for the momentum distribution, $f(p)p^4$, in a $M_{\rm s}=3.0$ shock, based on the test-particle DSA model.
The red line shows the injected population, $f_{\rm e,inj}(p)$, in Equation (\ref{feinj}).
The black dashed line shows a power-law spectrum of pre-existing fossil electrons, $f_{\rm e,foss}(p)$, with the slope, $s=4.7$, and the cutoff momentum,
$p_{\rm c}/m_{\rm e}c=10^3$.
The blue dot-dashed line show the spectrum of re-accelerated electrons, $f_{\rm e,reacc}(p)$ in Equation (\ref{freacc}).
The green vertical line denotes $p_{\rm ref}=Q_{\rm i,e} p_{\rm th,e}$ with $Q_{\rm i,e}=3.8$, above which suprathermal
electrons are reflected at the shock ramp and accelerated by Fermi-I acceleration.
Note that the amplitude of $f_{\rm e,reacc}(p)$ scales with the adopted normalization factor, $f_{\rm o}$, and so the relative importance between $f_{\rm e,reacc}$ and $f_{\rm e,inj}$ depends on it.
The proton spectrum, including both the postshock Maxwellian and injected DSA power-law components, is shown by the black solid line for comparison.
The magenta vertical line demarcates the injection momentum, $p_{\rm inj} = Q_{\rm i,p} p_{\rm th,p}$  with $Q_{\rm i,p}=3.8$, above which particles can
undergo the full DSA process across the shock transition.
\label{f1}
}
\end{figure}
\begin{figure*}[t]
\vskip -0.7cm
\hskip 0cm
\centerline{\includegraphics[width=0.95\textwidth]{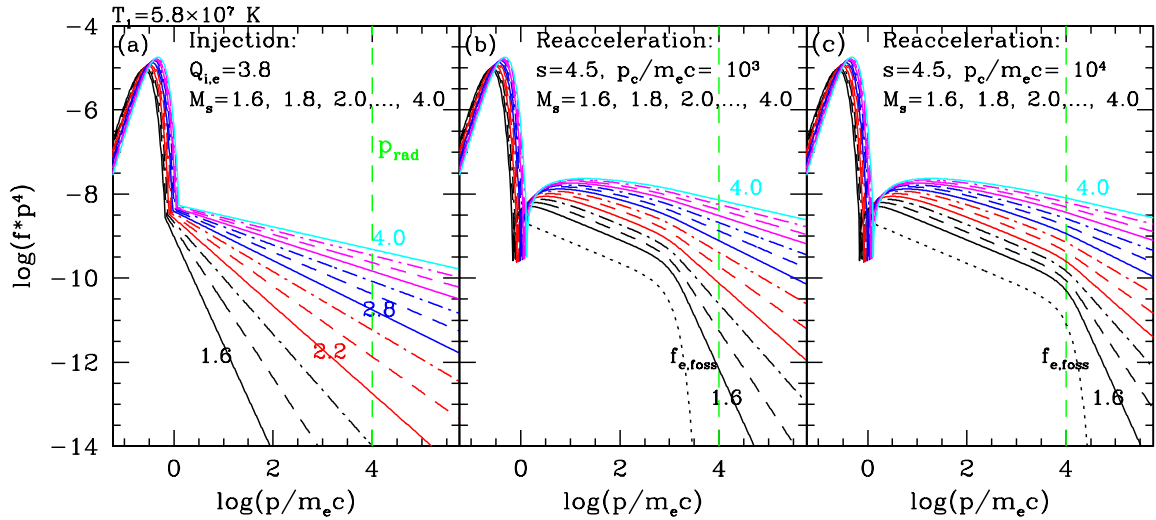}}
\vskip -9cm
\caption{Semi-analytic DSA model for $f(p)p^4$ for shocks with $M_s=1.6 - 4.0$ and $T_1=5.8\times10^7$~K:
(a) injected spectrum, $f_{\rm e,inj}(p)$, with $Q_{\rm i,e}=3.8$,
(b) re-accelerated spectrum, $f_{\rm e,reacc}(p)$, for the power-law spectrum of fossil electrons with $s=4.5$ and $p_{\rm c}/m_{\rm e}c=10^3$,
(c) the same as (b) except $p_{\rm c}/m_{\rm e}c=10^4$.
The blue dashed lines correspond to the models with $M_{\rm s}=3.0$ and $q=s=4.5$.
Note that the re-accelerated spectrum scales with the normalization factor of $f_{\rm e,foss}$, \ie $f_{\rm e,reacc}\propto f_{\rm o}$.
The green vertical lines denotes $p_{\rm rad}/m_{\rm e}c=10^4$.
\label{f2}
}
\end{figure*}

\subsection{Test-Particle Solutions for Injection-only Case}
\label{s2.2}

Here we adopt the test-particle solutions of DSA,
because dynamical feedbacks of CRp and CRe are expected be insignificant at weak ICM shocks \citep[e.g.,][]{ryu2019}.
Then the isotropic part of the momentum distribution function at the shock position
can be approximated by a power-law spectrum with a super-exponential cutoff.
For CRp spectrum,
\begin{equation}
f_{\rm p,inj}(p) \approx f_{\rm inj}\cdot \left(p \over p_{\rm inj} \right) ^{-q} \exp\left(-{p^2 \over p_{\rm max}^2} \right).
\label{fpinj}
\end{equation}
The normalization factor at $p_{\rm inj}$ is given as
\begin{equation}
f_{\rm inj} = {n_{\rm p,2} \over \pi^{1.5}} p_{\rm th,p}^{-3} \exp(-Q_{\rm i,p}^2),
\label{finj}
\end{equation}
where $n_{\rm p,2}$ is the postshock proton number density.
The maximum momentum of CRp achieved by the shock age of $t$ can be estimated as
\begin{equation}
{p_{\rm max} \over m_{\rm p} c } \approx {{\sigma -1} \over {6\sigma}} {u_{\rm s}^2 \over \kappa^*} t,
\end{equation}
where $\sigma=n_2/n_1$ is the shock compression ratio and $\kappa^*$ is the diffusion coefficient at
$p= m_{\rm p} c$ \citep{kang2011}. For ICM shocks, $p_{\rm max} / m_{\rm p} c \gg 1$, so the exponential cutoff at $p_{\rm max}$ is not important for weak shocks.

Similarly, the test-particle spectrum of CRe can be expressed as
\begin{equation}
f_{\rm e,inj}(p) \approx f_{\rm ref}\cdot \left(p \over p_{\rm ref} \right) ^{-q} \exp\left(-{p^2 \over p_{\rm eq}^2} \right).
\label{feinj}
\end{equation}
The normalization factor at $p_{\rm ref}$ is given as
\begin{equation}
f_{\rm ref} = {n_{\rm e,2} \over \pi^{1.5}} p_{\rm th,e}^{-3} \exp(-Q_{\rm i,e}^2),
\label{fref}
\end{equation}
where $n_{\rm e,2}$ is the postshock electron number density.
The cutoff momentum, $p_{\rm eq}$, can be derived from the equilibrium condition that the DSA momentum gains per cycle
are equal to the synchrotron/iC losses per cycle \citep{kang11}:
\begin{equation}
p_{\rm eq}= {m_e^2 c^2 u_s \over \sqrt{4e^3q/27}} \left({B_1 \over {B_{\rm e,1}^2 + B_{\rm e,2}^2}}\right)^{1/2},
\label{peq}
\end{equation}
where the `effective' magnetic field strength $B_{\rm e}^2= B^2 + B_{\rm rad}^2$ takes account for
radiative losses due to both synchrotron and iC processes,
where $B_{\rm rad}=3.24\muG(1+z)^2$ corresponds to the cosmic background
radiation at redshift $z$.
Here, we assume the Bohm diffusion for DSA, and set $z=0.2$ as a reference epoch and so $B_{\rm rad}=4.7\muG$.
For typical ICM shock parameters, it becomes
\begin{equation}
{p_{\rm eq}\over {m_{\rm e}c}} \approx {6.75\times 10^9 \over q^{1/2}} \left({u_s \over {1000 \kms}}\right) \left({B_1 \over {B_{\rm e,1}^2 + B_{\rm e,2}^2}}\right)^{1/2},
\label{peq2}
\end{equation}
where the magnetic field strength is expressed in units of $\mu G$.
Again, $p_{\rm eq}/m_{\rm e}c\gg 10^4$, so the exponential cutoff is not important for weak shocks.

With the DSA model spectra given in Equations (\ref{fpinj}) and (\ref{feinj}),
if $Q_{\rm i,p}= Q_{\rm i,e}$ as assumed here, then
the ratio of $f_{\rm p,inj}$ to $f_{\rm e,inj}$ at $p=p_{\rm inj}$ can be estimated as
\begin{equation}
K_{p/e}\equiv {f_{\rm p,inj}(p_{\rm inj}) \over f_{\rm e,inj}(p_{\rm inj})} = \left({p_{\rm th,p} \over p_{\rm th,e}} \right)^{q-3} = \left( {m_{\rm p} \over m_{\rm e}}\right) ^{(q-3)/2},
\end{equation}
where $K_{p/e}$ is equivalent to the CRp-to-CRe number ratio.
For example, in the case of a $M_{\rm s}=3.0$ shock with $q=4.5$, $K_{p/e}=280$,
but with a caveat that protons (electrons) are accelerated at $Q_\parallel$ ($Q_\perp$) shocks.

In Figure \ref{f1}, we illustrate the thermal Maxwellian distribution and the test-particle power-law spectrum, $f_{\rm p,inj}(p)$,
for protons, which are demarcated by the magenta line of $p_{\rm inj}$. 
The shock parameters adopted here are $Q_{i,p}= 3.8$,  $M_s=3$ and $T_1=5.8\times 10^7$~K, and $q=4.5$.
Also, the thermal Maxwellian distribution and $f_{\rm e,inj}(p)$ for electrons are demarcated 
by the green line of $p_{\rm ref}$ with $Q_{i,e}= 3.8$. 
This clearly demonstrates that, in order to get injected to DSA, 
the reflected electrons need to be energized by a factor of 
$p_{\rm inj}/p_{\rm ref}\approx \sqrt{m_{\rm p}/m_{\rm e}}$ or so.

\begin{figure*}[t]
\vskip -0.7cm
\hskip -0.1cm
\centerline{\includegraphics[width=0.90\textwidth]{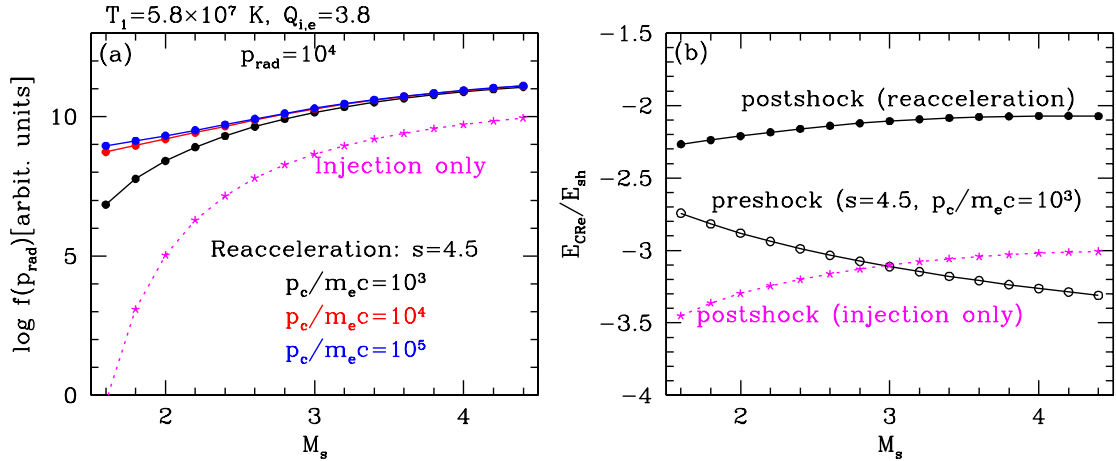}}
\vskip -9.2 cm
\caption{
(a) Amplitude of $f(p_{\rm rad})$ at $p_{\rm rad}=10^4 m_{\rm e}c$ for the re-acceleration case
with the slope, $s=4.5$, and the cutoff momentum, $p/m_{\rm e}c=10^3$ (black circles), $10^4$ (red circles), and $10^4$ (blue circles).
The magenta line with asterisks shows $f(p_{\rm rad})$ for the injection-only case.
The semi-analytic DSA spectra ($Q_{\rm i,e}=3.8$) shown in Figure \ref{f2} are used.
(b) CRe pressure, $E_{\rm CRe}$, in units of $E_{\rm sh}=\rho_1 u_s^2/2$.
The preshock $E_{\rm CRe,1}$ due to fossil CRe with $s=4.5$ and $p_{\rm c}/m_{\rm e}c=10^3$ is shown by open circles,
while the postshock $E_{\rm CRe,2}$ due to re-accelerated CRe is shown by closed circles.
The magenta line with asterisks shows $E_{\rm CRe,2}$ for the injection-only case.
Again, the semi-analytic DSA spectra in Figure \ref{f2} are used.
\label{f3}
}
\end{figure*}

\begin{figure*}[t]
\vskip -0.7cm
\hskip 0cm
\centerline{\includegraphics[width=0.95\textwidth]{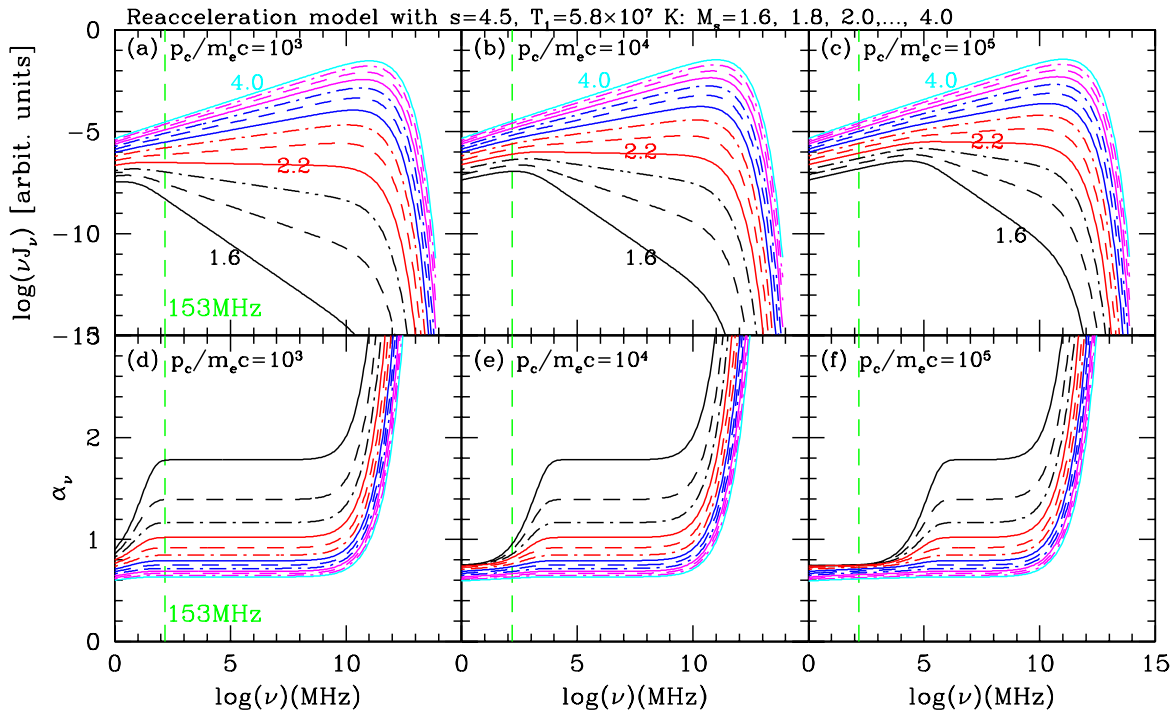}}
\vskip -6.5 cm
\caption{
Upper panels: Synchrotron spectrum, $\nu j_{\nu}$, due to $f_{\rm e,reacc}(p)$ re-accelerated at shock 
with $M_s=1.6 - 4.0$ and $T_1=5.8\times10^7$~K in the presence of
fossil CRe with the slope, $s=4.5$, and the cutoff momentum, (a) $p/m_{\rm e}c=10^3$, (b) $p/m_{\rm e}c=10^4$, (c) $p/m_{\rm e}c=10^5$.
Note that the results for the {\it in situ} acceleration model
are not shown, because the corresponding synchrotron spectrum is a simple power-laws with a cutoff.
The semi-analytic DSA spectra ($Q_{\rm i,e}=3.8$) shown in Figure \ref{f2} are used.
The emissivity $\nu j_{\nu}$ is plotted in arbitrary units.
The blue dashed lines correspond to the models with $M_{\rm s}=3.0$ and $q=s=4.5$.
The green vertical denotes $\nu=153$~MHz. 
Lower panels: Synchrotron spectral index, $\alpha_{\nu}= -d \ln j_{\nu}/d \ln \nu $,
for the radiation spectra shown in the upper panels.
\label{f4}
}
\end{figure*}

\begin{figure*}[t]
\vskip -0.7cm
\hskip 0cm
\centerline{\includegraphics[width=0.90\textwidth]{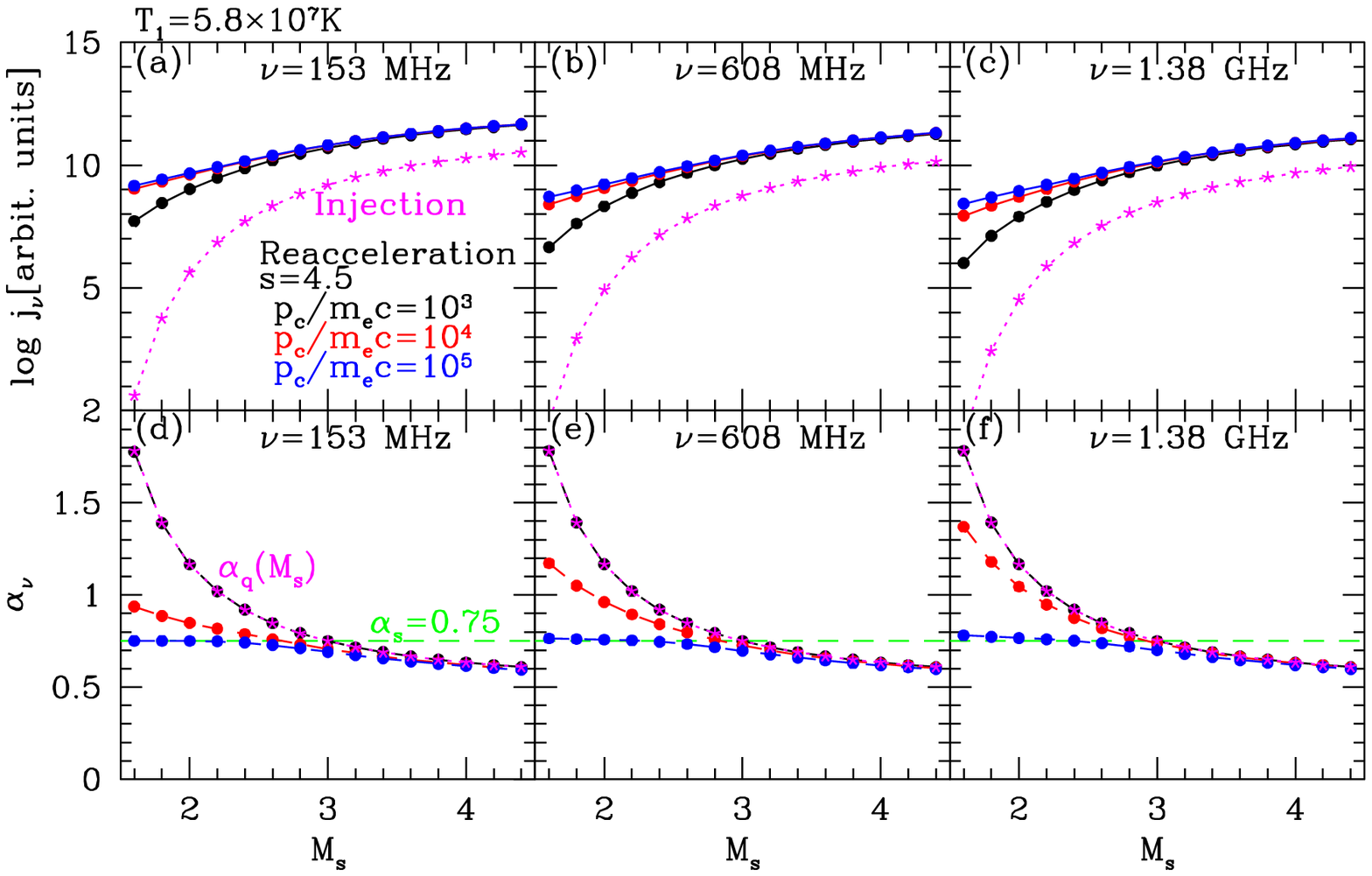}}
\vskip -6.0 cm
\caption{
Upper panels: Amplitudes of $j_{\nu}$ at $153$~MHz, 608~MHz, and 1.38GHz for the synchrotron spectra shown in Figure \ref{f4}.
Three models with fossil CRe with the slope, $s=4.5$, and the cutoff momentum, $p/m_{\rm e}c=10^3$ (black circles), $10^4$ (red circles), and $10^5$ (blue circles) are shown.
The magenta lines with asterisks show the same quantities for the injection-only case.
Note that $j_{\nu}$ scales with the adopted normalization factor, $f_{\rm o}$ and is plotted in arbitrary units here.
Lower panels: Synchrotron spectral index, $\alpha_{\nu}= -d \ln j_{\nu}/d \ln \nu $, for the same models shown in the upper panels.
Note that the black line (re-acceleration case with $p/m_{\rm e}c=10^3$) coincides with the magenta line (injection-only case), which also
corresponds to $\alpha_{\rm q}(M_{\rm s}) = (q-3)/2$.
The green horizontal lines denote $\alpha_{\rm s} = (s-3)/2=0.75$.
\label{f5}
}
\end{figure*}

\subsection{Re-acceleration of Fossil CR Electrons}
\label{s2.3}

For the preshock population of fossil CRe, we adopt a power-law spectrum with the slope $s$ and a cutoff momentum $p_c$ for $p>p_{\rm ref}$: 
\begin{equation}
f_{\rm e,foss}(p) = f_{\rm o} \cdot \left(p \over p_{\rm ref} \right) ^{-s} \exp\left(-{p^2 \over p_{\rm c}^2} \right),
\label{fpre}
\end{equation}
where $p_{\rm c}/ m_{\rm e} c=10^3-10^5$ is considered in this discussion, 
and the normalization factor, $f_{\rm o}$, determines the amount of fossil CRe.
Then, the re-accelerated population at the shock can be calculated semi-analytically by the following integration:
\begin{equation}
f_{\rm e,reacc}(p)= q \cdot p^{-q} \int_{p_{\rm ref}}^p p^{\prime q-1} f_{\rm e,foss} (p^\prime) dp^\prime 
\label{freacc}
\end{equation}
\citep{dru1983,kang2011}.
In Figure \ref{f1}, we show an example of $f_{\rm e,foss}(p)$ with $s=4.7$, and  $p_{\rm c}/ m_{\rm e}=10^3$
in the black dashed line, while its re-accelerated spectrum, $f_{\rm e,reacc}(p)$, at a $M_{\rm s}=3$ shock is shown in the blue dot-dashed line.
Note that the normalization factor is set as $f_{\rm o}=f_{\rm ref}$ in Equation (\ref{fref})
with $Q_{\rm i,e}=3.8$ just for illustration purpose only.

For a power-law fossil population without a cutoff, $f_{\rm e,foss}\propto  p^{-s}$, 
the re-accelerated spectrum can be obtained by direct integration \citep{kang2011}: for $p\ge p_{\rm ref}$
\begin{eqnarray}
f_{\rm e,reacc}(p) = \left\{ \begin{array}{rl} 
{q\over(q-s)} \left[ 1-({p / p_{\rm ref}})^{-q+s}\right] f_{\rm e,foss}(p),\\
\mbox{ if $q\neq s$};\\ 
q \ln(p/p_{\rm ref}) f_{\rm e,foss}(p),\\
\mbox{ if $q= s$}.
\end{array} \right.
\label{f2p}
\end{eqnarray}

Although we do not explicitly show it here,
the re-accelerated spectrum of pre-existing protons, $f_{\rm p,reacc}(p)$, can be described by the same integration as Equation (\ref{freacc}), 
except that the lower bound should be replaced with $p_{\rm inj}$ and $f_{\rm e,foss}(p)$ should be replaced
with a pre-existing proton population, $f_{\rm p,pre}(p)$, with appropriate parameters, $s$, $p_{\rm c}$,
and $f_{\rm o}$.

\section{Application to Radio Relics}
\label{s3}

\subsection{DSA Model Spectrum}
\label{s3.1}

We apply the DSA models given in Equations (\ref{feinj}) and (\ref{freacc}) to calculate the energy spectrum of accelerated electrons at weak shocks propagating into the preshock gas with $T_1=5.8\times10^7$~K.
Panel (a) of Figure \ref{f2} shows the injection spectrum, $f_{\rm e,inj}(p)$, for shocks with 
$M_{\rm s}=1.6-4.0$ (with the increment $\Delta M_{\rm s} = 0.2$).  
Considering that the synchrotron emission from mono-energetic electrons with the Lorentz factor, 
$\gamma_{\rm e}$, peaks around the characteristic frequency,
$\nu_{\rm peak}\approx 130{\rm MHz}({\gamma_e / {10^4}})^{2} (B/ 1\muG)$,
we compare the amplitude of $f_{\rm e,inj}(p_{\rm rad})$, where $p_{\rm rad}= 10^4m_{\rm e}c$, by the green vertical dashed line.
Then, panel (a) of Figure \ref{f3} illustrates how $f_{\rm e,inj}(p_{\rm rad})$ depends on the shock Mach number by the magenta dotted line.
In the case of the {\it in situ} acceleration model, $f_{\rm e,inj}(p_{\rm rad})$ increases by a factor of $4.2\times 10^3$ for the 
range of $M_{\rm s}=2.0 - 3.0$.
This strong dependence is even stronger at lower Mach number, so that $f_{\rm e,inj}(p_{\rm rad})$ decreases almost by a factor of $90$, when
$M_s$ decreases from 2.0 to 1.8.
This implies that the radio surface brightness could vary extremely sensitively with $M_s$, 
when a radio relic consists of multiple shocks with slightly different Mach numbers \citep{roh2019}.

Panels (b) and (c) of Figure \ref{f2} show the re-accelerated spectra, $f_{\rm e,reacc}(p)$, of 
the fossil electron spectrum, $f_{\rm e,foss}(p)$, with $s=4.5$ and $p_{\rm c}/m_{\rm e}c=10^3$
and $10^4$, respectively.
Again, we set $f_{\rm o}=f_{\rm ref}$ with $Q_{\rm i,e}=3.8$ as in Figure \ref{f1}.
For stronger shocks with $M_{\rm s} \ge 3.0$, the re-accelerated spectrum is flatter than the fossil spectrum, 
\ie, $q\le s$. Hence fossil CRe serve only as seed particles, and so $f_{\rm e,reacc}(p)$ does not 
depend on the cutoff momentum, $p_{\rm c}$.
For weaker shocks with $M_{\rm s} < 3.0$, on the other hand, $q>s$. 
Hence, $f_{\rm e,reacc}(p)$ depends on $p_{\rm c}$ for $p> p_{\rm c}$, as can be seen in Figure \ref{f2}.
Thus, in the case of weaker shocks with $q>s$, the cutoff $p_{\rm c}$ should be high enough 
in order to obtain the simple power-law, $f_{\rm e,reacc}(p)\propto p^{-s}$ for $p>p_{\rm rad}$.
Again, panel (a) of Figure \ref{f3} shows how $f_{\rm e,reacc}(p_{\rm rad})$ for the re-acceleration model 
depends on $M_{\rm s}$ for the three models of $p_{\rm c}/m_{\rm e}c=10^3$ (black), $10^4$ (red), and $10^5$ (blue).
It increases only by a factor of $55$ for the range of $M_{\rm s}=2.0 - 3.0$ for the models with $p_{\rm c}/m_{\rm e}c=10^3$.
In the case of $p_{\rm c}/m_{\rm e}c=10^5$, it increases by an even smaller factor, $9.6$, for $M_{\rm s}=2.0 - 3.0$.
Thus, the amplitude of $f_{\rm e,reacc}(p_{\rm rad})$ of radio-emitting electrons for the re-acceleration model has a much weaker dependence on $M_{\rm s}$, compared to $f_{\rm e,inj}(p_{\rm rad})$ for the {\it in situ} acceleration model.
Moreover, this dependence becomes weaker if the cutoff momentum is higher than 
$p_{\rm c}/m_{\rm e}c \gtrsim 10^5$.

Panel (b) of Figure \ref{f3} shows the ratio of $E_{\rm CRe}/E_{\rm sh}$ due to the following three CRe spectra:
(1) $f_{\rm e,reacc}(p)$ in the postshock region,
(2) $f_{\rm e,foss}(p)$ with $p_{\rm c}/m_{\rm e}c=10^3$ in the preshock region, and
(3) $f_{\rm e,inj}(p)$ in the postshock region.
With the adopted value of $Q_{\rm i,e}=3.8$, the {\it in situ} acceleration model (magenta asterisks)
predicts that
$E_{\rm CRe,2}/E_{\rm sh}\approx 10^{-3.5}- 10^{-3}$ for $1.6 \lesssim M_{\rm s} \lesssim 4.4$.
In the case of the re-acceleration model with $s=4.5$, $p_{\rm c}/m_{\rm e}c=10^3$, and 
$f_{\rm o}=f_{\rm ref}$, the postshock ratio (closed circles) varies rather slowly, $E_{\rm CRe,2}/E_{\rm sh}\approx 10^{-2.3}- 10^{-2.1}$. 
Of course, this ratio is arbitrary here and scales with the adopted normalization factor, $f_{\rm o}$, of the fossil CRe population.
Note that the preshock ratio (open circles), $E_{\rm CRe,1}/E_{\rm sh}$, decreases with increasing $M_{\rm s}$, because the denominator increases with $M_{\rm s}^2$.

\subsection{Radio Synchrotron Emission}
\label{s3.2}

We then calculate the synchrotron volume emissivity, $j_{\nu}(\nu)$, due to $f_{\rm e,reacc}(p)$
shown in Figure \ref{f2},
with the postshock magnetic field strength, $B_2=B_1 \sqrt{1/3+2\sigma^2/3}$ (where $B_1= \mu G$),
in order to illustrate how the radio spectrum changes with the shock Mach number.
We do not show explicitly the emissivity spectrum for the {\it in situ} acceleration case, 
since both $f_{\rm e,inj}(p)$ and $j_{\nu}(\nu)$ are simple power-laws with an exponential cutoff.

The top panels of Figure \ref{f4} shows $\nu j_{\nu}$, while the bottom panel shows its spectral index,
$\alpha_{\nu}= -d \ln j_{\nu}/d \ln \nu $.
Note that $j_{\nu}$ scales with the adopted normalization factor, $f_{\rm o}$ and is plotted in arbitrary units here.
For stronger shocks with $q<s$, $j_{\nu}(\nu)$ is a power-law with $\alpha_{\rm q}= (q-3)/2={{(M_s^2+3)}/{2(M_s^2-1)}}$ with a cutoff.
For weaker shocks with $q>s$, on the other hand, the radio spectrum depends on the cutoff momentum of the fossil CRe spectrum, 
$p_{\rm c}$, as well as $M_{\rm s}$, as expected obviously from $f_{\rm e,reacc}(p)$ in Figure \ref{f2}.
At these weaker shocks, the slope, $\alpha_{\nu}$, gradually increases from $(s-3)/2$ to $(q-3)/2$, as the frequency increases. 
Hence, the fossil CRe power-law should extend to well above $p_{\rm c}/m_{\rm e}c \gtrsim 10^5$,
in order for the spectral index to be determined by the slope of fossil CRe, \ie $\alpha_{\rm s}= (s-3)/2$,
for $\nu\lesssim 10$~GHz (see panel (f) of Figure \ref{f4}).
For example, if the power-law spectrum of fossil CRe extends only up to $p_{\rm c}/m_{\rm e}c \lesssim 10^3$,
the radio spectral index due to postshock CRe becomes $(q-3)/2$ for
$\nu\gtrsim 153$~MHz (see panel (d) of Figure \ref{f4}).

Figure \ref{f5} shows the relative values of $j_{\nu}$ and $\alpha_{\nu}$
at three typical observation frequencies, $\nu_{\rm obs}=153$~MHz, 608~MHz, and 1.38~GHz \citep[e.g.,][]{vanweeren10}.
Similarly to the case of $f(p_{\rm rad})$, both $j_{\nu}(\nu_{\rm obs})$ and $\alpha_{\nu}(\nu_{\rm obs})$
vary strongly with $M_{\rm s}$ for the {\it in situ} acceleration model (magenta asterisks).
For example, $j_{153{\rm MHz}}$ increases by a factor of $3.7\times10^3$,
$j_{608{\rm MHz}}$ by a factor of $6.5\times10^3$,
and $j_{1.38{\rm GHz}}$ by a factor of $9.2\times10^3$, as $M_{\rm s}$ increases from $2.0$ to $3.0$.
So the Mach number dependence is a bit stronger at higher observational frequencies.
Again, the re-acceleration models (filled circles) exhibit much weaker dependence on $M_{\rm s}$.
For the model with $p_{\rm c}/m_{\rm e}c = 10^3$ (black filled circles), $j_{153{\rm MHz}}$ increases by a factor of 48, and $j_{1.38{\rm GHz}}$ by a factor of 120 for $M_{\rm s}=2.0-3.0$.
For the model with $p_{\rm c}/m_{\rm e}c = 10^5$ (blue filled circles), on the other hand,
$j_{153{\rm MHz}}$ increases only by a factor of 14, and $j_{1.38{\rm GHz}}$ by a factor of 16 
for $M_{\rm s}=2.0-3.0$.

For stronger shocks with $q\le s$, even the re-acceleration models
have the spectral index that follows the injection index, $\alpha_{\rm q}$ (magenta asterisks).
So in the bottom panels of Figure \ref{f5}, all symbols (black, red, blue, and magenta) overlap with each other
for $M_{\rm s} \ge 3.0$.
For weaker models with $q>s$, again the spectral index depends on $p_{\rm c}$.
For the models with $p_{\rm c}/m_{\rm e}c = 10^5$ (blue filled circles), $\alpha_{\nu}\approx \alpha_{\rm s}$
for $M_{\rm s}<3$.
In the case of the models $p_{\rm c}/m_{\rm e}c = 10^3$ (black filled circles), the fossil CRe serve as 
only seed particles, so the spectral indices at the three observational frequencies becomes the same as 
the injection index, $\alpha_{\rm q} (M_{\rm s})$.

In conclusion, if a radio relic is composed of multiple shocks with slightly different Mach numbers \citep{roh2019},
the surface brightness fluctuations could be much larger in the {\it in situ} acceleration model,
compared to the re-acceleration model. 
But the variations in the spectral index profile should be much smaller.
Relatively smooth profiles of radio flux along the edge of some observed radio relics, 
such as the Sausage relic \citep{hoang17} and the Toothbrush relic \citep{vanweeren16}, 
probably indicate that the re-acceleration might play a significant role there.

\section{Summary}
\label{s4}

Based on the recent studies using plasma kinetic simulations \citep[e.g.,][]{guo14,matsukiyo15,park2015, kang2019,trotta2019,kobzar2019},
here we suggest semi-analytic DSA models for the electron (re)-acceleration at weak $Q_\perp$-shocks 
in the test-particle regime.
They rely on the following working assumptions:
(1) at $Q_\perp$-shocks of all Mach numbers in the test-particle regime (\ie $M_{\rm s}\lesssim 4$), 
electrons can be pre-accelerated from thermal pool by both electron and ion kinetic instabilities 
and injected to the DAS process, and
(2) the momentum distribution function of (re)-accelerated electrons follows the prediction of the DSA theory for 
$p\ge p_{\rm ref}= Q_{\rm i,e} p_{\rm th,e}$ with $Q_{\rm i,e} \approx 3.5-3.8$.
However, it remains uncertain if and how subcritical shocks with $M_{\rm s} \lesssim 2.3$ could inject electrons to the DSA process \citep{kang2019} or
re-accelerate pre-existing fossil CR electrons through DSA.
We include the electron (re)-acceleration in subcritical shocks here, because, 
in some of observed radio relics,
the shock Mach number is estimated to be less than 2.3 \citep[e.g.,][]{vanweeren16}.

Then, the momentum distribution of accelerated electrons can be represented by the simple power-law with a cutoff
given in Equation (\ref{feinj}) for the {\it in situ} acceleration model.
For the re-acceleration model with fossil CRe, which is specified by the three parameters,
the slope, $s$, the cutoff, $p_{\rm c}$, and the normalization factor, $f_{\rm o}$, 
the re-accelerated spectrum can be integrated semi-analytically as in Equation (\ref{freacc}).

We explore how our model spectrum of CRe varies with the parameters such as
$M_{\rm s}$, $s$, and $p_{\rm c}$ in the case of weak shocks with $M_{\rm s}=1.6-4.4$
for the two types of DSA models: the {\it in situ} acceleration model and the re-acceleration model.
The main findings can be summarized as follows:

\begin{enumerate}

\item For stronger shocks with $q = 4M_{\rm s}^2/(M_{\rm s}^2-1) \le s$,
the re-accelerated spectrum becomes a power-law, $f_{\rm e,reacc}(p)\propto p^{-q}$,
and it does not depend on $p_{\rm c}$.
The radio synchrotron spectrum becomes also a power-law with $\alpha_{\nu} \approx \alpha_{\rm q}= (q-3)/2={{(M_s^2+3)}/{2(M_s^2-1)}}$ and an appropriate cutoff.

\item For weaker shocks with $q>s$, on the other hand, $f_{\rm e,reacc}(p)$ depends on the cutoff $p_{\rm c}$. 
Only for $p_{\rm c}/m_{\rm e}c \gtrsim 10^5$,
the radio synchrotron spectrum has the spectral index,
$\alpha_{\nu} \approx \alpha_{\rm s} =(s-3)/2$ for the observation frequencies in the range of $\nu_{\rm obs} \approx 100$~MHz $-10$~GHz.

\item If $p_{\rm c}/m_{\rm e}c \lesssim 10^3$, the fossil CRe provide only seed particles to DSA, and hence
the spectral index is similar to the injection index, $\alpha_{\nu}\approx \alpha_{\rm q}$.

\item In the {\it in situ} acceleration model, the radio synchrotron emissivity, $j_{\nu}$, depends strongly 
on $M_{\rm s}$, and it increases by a factor of $10^3-10^4$, as $M_{\rm s}$ increases from 2.0 to 3.0. 
But it varies by a factor of only $15$ or so in the re-acceleration model with 
$p_{\rm c}/m_{\rm e}c = 10^5$ for the same range of $M_{\rm s}$.
In the case of a lower cutoff at $p_{\rm c}/m_{\rm e}c = 10^3$, $j_{\rm 153MHZ}$ increases by a factor of 
48, and  $j_{\rm 1.38GHZ}$ by a factor of 120 for the same range of $M_{\rm s}$.

\end{enumerate}

Considering that the spatial profiles of radio flux and spectral index vary rather smoothly along the edge 
of some observed radio relics \citep[e.g.,][]{vanweeren16,hoang17}, and that giant radio relics on Mpc scales are likely to consist of
multiple shocks with different $M_{\rm s}$ \citep[e.g.,][]{roh2019}, our results imply that the re-acceleration of fossil CRe
is important in understanding the origin of radio relics.

\acknowledgments

This work was supported by a 2-Year Research Grant of Pusan National University.
The author thanks D. Ryu for stimulating discussions at the initial stage of this work.


\end{document}